# A physics-aware deep learning model for energy localization in multiscale shock-to-detonation simulations of heterogeneous energetic materials

Phong C.H Nguyen[1], Yen-Thi Nguyen[2], Pradeep K. Seshadri[2], Joseph B. Choi[1], H.S Udaykumar[*,2], and Stephen Baek[*,1,3]

[1]University of Virginia, School of Data Science, Charlottesville, VA 22904, United States
[2]University of Iowa, Department of Mechanical Engineering, Iowa City, IA 52242, United States
[3]University of Virginia, Department of Mechanical and Aerospace Engineering, Charlottesville, VA 22903, United States

**Abstract:** Predictive simulations of the shock-to-detonation transition (SDT) in heterogeneous energetic materials (EM) are vital to the design and control of their energy release and sensitivity. Due to the complexity of the thermo-mechanics of EM during the SDT, both macro-scale response and sub-grid mesoscale energy localization must be captured accurately. This work proposes an efficient and accurate multiscale framework for SDT simulations of EM. We introduce a new approach for SDT simulation by using deep learning to model the mesoscale energy localization of shock-initiated EM microstructures. The proposed multiscale modeling framework is divided into two stages. First, a physics-aware recurrent convolutional neural network (PARC) is used to model the mesoscale energy localization of shock-initiated heterogeneous EM microstructures. PARC is trained using direct numerical simulations (DNS) of hotspot ignition and growth within microstructures of pressed HMX material subjected to different input shock strengths. After training, PARC is employed to supply hotspot ignition and growth rates for macroscale SDT simulations. We show that PARC can play the role of a surrogate model in a multiscale simulation framework, while drastically reducing the computation cost and providing improved representations of the sub-grid physics. The proposed multiscale modeling approach will provide a new tool for material scientists in designing high-performance and safer energetic materials.

## 1 Introduction

Heterogeneous energetic materials (EM), which are composites of organic crystals, plasticizers, metals, and other inclusions, are the key component in various military and civilian applications. Understanding the shock-to-detonation transition (SDT) in EM is important to control energy release and for the safe handling of propellant and explosive materials. SDT in EM is initiated due to energy localization at the mesoscale, i.e., at the scale of grains and defects in their microstructures. When a shock propagates through EM microstructures, energy localization occurs due to the presence of voids, cracks, and interfaces, leading to the creation of regions with high temperature called hotspots. With a sufficient number of generated hotspots, the chemical energy release will be rapid enough to couple with and strengthen the imposed shock wave, initiating a detonation [1–3]. Due to the strong connection between the meso-scale energy localization and the SDT of EM, the meso-scale thermo-mechanics simulation are needed to predict the macroscale response of EM accurately.

Currently, it is nearly impossible to capture the localized mesoscale hotspot formation in macroscopic samples due to computational issues. Resolving mesoscale hotspot physics requires a stringent spatial and temporal resolution, resulting in a computational model with an enormous number of grid points. Such heavy computation models are costly, laborious, and non-scalable. A solution to this problem is applying multiscale approaches [4,5]. In a multiscale setting, mesoscale physics such as hotspot initiation and growth are conveyed to the macroscale as a sub-grid source or closure term as opposed to being solved in a fully resolved model with detailed features. Therefore, the computational cost for modeling the SDT of EM can be reduced and the EM design process can be facilitated.

This work presents a multiscale framework that can model the SDT of EM efficiently with a multi-scale representation of the hotspot evolution. The key contribution of the proposed multiscale framework is the use of a deep-learning model, namely physics-aware recurrent convolution (PARC) neural networks, proposed by Nguyen et al. [6], to supply the subgrid (meso-scale) reaction progress rate of shocked EM to the macro-scale computation. Once trained, PARC can rapidly predict the temperature and pressure field evolutions of shock-initiated EM microstructures under a given applied pressure. As demonstrated previously by Nguyen et al. [6], the formation and growth of hotspots predicted by PARC had a comparable accuracy to that of direct numerical simulation (DNS).

Furthermore, PARC provides multiple orders of magnitude reduction in computational time relative to full-fledged DNS—from hours on a high-performance computing cluster (HPC) in the case of DNS to less than a second on a commodity laptop with PARC. Utilizing the rapid hotspot evolution prediction capability of PARC, this paper demonstrates that the reaction progress rates of shocked EM of a given microstructure can be used to inform macroscale SDT simulations. We first extend the PARC architecture to allow it to make predictions with different applied shock strengths. We then integrate PARC into a multiscale framework for SDT simulation.

The rest of the paper is organized as follows. Section 2 reviews the state-of-the-art in multiscale SDT simulations of EM. The method section (Section 3) presents the improved PARC model and discusses how the hotspot physics analysis at the mesoscale can be used to inform the SDT simulation at the macroscale. Section 4 describes the validation of the improved PARC model and its application in multiscale SDT simulations. Finally, conclusions and remarks are given in Section 5.

## 2 Background

Multiscale simulations of energetic materials have been drawing attention due to the microstructure-aware nature of such approaches in modeling the complex physics of SDT in EM. In a typical multiscale SDT framework, the reactive meso-mechanics of EM is modeled at two distinct scales, viz., macroscale and mesoscale. At the macroscale, detailed features, including voids, cracks, grains, and interfaces, are unresolved and the material is treated as homogeneous. The EM at this scale can be considered as single-phase [5,7,8] or multi-phase mixtures [9]. In the mesoscale analysis, the detailed thermo-mechanics of energy localization, including hotspot formation, growth and interaction due to the heterogeneity of EM microstructures, will be captured with fully resolved, high-fidelity models. The communication between the two-scale physics must be performed through physically correct closure models. Such closure models are used in the macroscale governing equation to describe the behavior of the homogeneous energetic material during the SDT [5].

The key quantity of interest (QoI) for the SDT simulations of EM is the *reaction progress variable* $\lambda$ [5,7,8,10]. The progress variable is often used as the indicator to determine the state of EM during the SDT process [4]. When $\lambda = 0$, the material is considered being unreacted; meanwhile, $\lambda = 1$ indicates a complete reaction. The *rate of reaction progress variable* $\dot{\lambda}$ governs the rate of energy release during the SDT and is used in the macroscale governing equation as a source term, either directly [8,11–13] or indirectly [14,15]. Common approaches to derive the reaction progress rate rely on the DNS of resolved material microstructures at the mesoscale [16–21]. However, such a computation is intensive and thus discourages the execution of concurrent multiscale modeling in which the mesoscale reaction progress rate is derived simultaneously or "on-line" with the macroscale simulation. To this end, sequential hierarchical multiscale approaches can be the potential solution and have been explored in recent years. In this approach, the mesoscale physics is simulated separately, or "off-line," from the macroscale physics. Consequently, surrogate models are constructed from mesoscale analysis results and embedded in macroscale governing equations as closure models. With a well-trained surrogate model, the repeated heavy computation of the fully resolved mesoscale energy localization model can be avoided while the SDT prediction accuracy is still maintained.

Following this direction, many works have been devoted to developing surrogate models for estimating the reaction progress rate of shocked EM microstructures and using them in a multiscale SDT simulation framework. For instance, Sen et al. (2018) [5] developed a meso-informed ignition and growth (MES-IG) model to establish the structure-property-performance (SPP) linkage using an ignition-and-growth framework [10,11]. MES-IG modeled the reaction progress rate as a function of applied pressure, density and microstructural parameters, including void diameter, void aspect ratio, and void orientation. In another work, Perry et al. (2018) [22] developed the physically-informed scaled uniform reactive flow PiSURF model in an attempt to achieve the same goal but using the scaled uniform reactive flow (SURF) [8] route. Despite achieving certain successes, in most of previous works, the effects of complex microstructural morphologies were not fully explored due to the simplified assumption of microstructural geometry and loading conditions [4,20,21,23,24]. The utilization of idealized and canonical microstructures models carries the advantages of simplifying the calculation and helps reduce the dimensionality of the design search space. However, such an approach is usually oversimplified and does not resemble the real behavior of shocked EM microstructures [17]. First, real EM microstructures contain various features that have strong impacts on energy localization (e.g., large and elongated cracks or tortuous voids). Such morphological complexity cannot be well captured with idealized geometry representations. Second, simplified energy localization models cannot adequately capture the interaction among (contorted or branched) voids. Thus, the detailed information on the ignition or the combination of multiple hotspots that strongly contributes to the detonation initiation is not properly modeled [18]. In summary, with idealized EM microstructure representations, the mesoscale information transmitted to macroscale simulations is not expressive enough and may lead to a decrease in the accuracy of the macroscale SDT simulation.

To address this concern, previous works examined and modeled the impact of non-ideal shapes and void-void interactions on the mesoscale energy localization of shocked EM microstructures [16–18]. Despite of emerging capabilities in capturing the detailed thermo-mechanics of shocked heterogeneous EM, simulations with non-idealized voids are computationally intensive [17,18]. A single, well-resolved mesoscale simulation can take hours to days on HPC clusters [6]. Moreover, the complexity and stochasticity of EM microstructure morphology requires a large morphological parameter space to be examined to capture the statistically representative reaction progress rate. These computational demands lead to a formidable cost for analyzing and constructing surrogate models and thus, delay the overall design process.

Recently, Nguyen et al. [6] proposed a deep-learning framework called PARC to assimilate the hotspot thermo-mechanics of shocked EM microstructures. In PARC, complex morphological characteristics of EM microstructures and their effects on hotspot formation were modeled and learned using recurrent convolutional neural networks (CNN). PARC modeled the governing differential equation of the time-evolving temperature and pressure fields within shocked EM microstructures using a recurrent CNN. The CNN-modeled governing equation was then solved via data-driven integration, modeled as another CNN. As such, PARC implemented a physics-aware architecture by mimicking how the actual thermo-mechanics problems are solved, but in a data-driven approach. The validation studies by Nguyen et al. [6] showed that PARC predicted QoIs agree well with those obtained using DNS. Enabled by a physics-aware architecture, PARC predicted the hotspot ignition and growth of shocked EM microstructures with an accuracy comparable to the ground truth (i.e., DNS) while requiring multiple orders lower computational time.

Motivated by this predictive capability and the significant improvement in computational efficiency, here we examine whether PARC can replace DNS for the mesoscale energy localization analysis upon which the reaction progress rate can be derived to inform the macroscale simulation model. In the present work, we attempt to integrate PARC into a multiscale SDT simulation framework to predict the critical energy of initiation of pressed HMX, a typical type of heterogeneous EM, subjected to different shock strengths. We demonstrate that, with this modeling capability enabled by PARC, we can achieve predictions for a typical heterogeneous EM that reproduces experimental data.

# 3 Methods

## 3.1 Multiscale modeling framework for shock-to-detonation simulation

Figure 1 illustrates the proposed multiscale SDT simulation framework for EM. First, resolved mesoscale simulations are performed to assimilate the reactive shock response of EM microstructures acquired either from scanning electron microscopes (SEM) [18] or synthesized using deep generative models [25] (Fig. 1(a-c)). The simulations were performed using the previously developed and well-tested numerical simulation code, SCIMITAR3D [19,26–28]. The simulations provide snapshots describing the evolution of temperature and pressure fields through the duration of the passage shock and the formation and growth of hotspots. The deep-learning model (Fig 1d), i.e., PARC, is trained on a set of SCIMITAR3D simulations. Once trained, PARC can accurately assimilate the hotspot ignition and growth within shocked EM microstructures and can account for the influence of microstructural morphology on the hotspot thermo-mechanics, such as collapses of voids and cracks.

After training, PARC is used as a sub-grid model to supply the reaction progress rate within EM microstructures to the homogenized macro-scale reactive dynamics solvers. Particularly, from PARC-predicted temperature and pressure field evolution (Fig. 1d), the rate of hotspot total area evolution over time, which represents the reaction progress rate, can be extracted to inform the macroscale simulation (Fig. 1e). At the macroscale (Fig 1f), the domain is tessellated into discrete control volumes that are considered homogeneous. Homogenized control volumes are supplied with an effective reaction rate surrogate model, which is taken to be a function of pressure and the time variable. Here, the sample-average reaction progress rate in response to different shock strengths computed by PARC are used to train the scale bridging surrogate model. With the reaction progress rate provided by PARC, the macroscale SDT simulation is performed, and the run-to-detonation distance can be measured.

The following sub-sections discuss in more detail all the components of the proposed framework. Section 3.2 introduces the enhanced PARC for the energy localization modeling in EM microstructures subjected to different shock strengths. Section 3.3 describes the derivation of the reaction progress rate from PARC-predicted temperature and pressure field evolutions. Finally, the macroscale SDT simulation with the reaction progress rate derived from PARC predictions is introduced in Section 3.4.

## 3.2 PARC for mesoscale hotspot ignition and growth modeling

### 3.2.1 PARC architecture for hotspot ignition and growth modeling with various shock strengths

PARC was designed to solve the governing differential equation of temperature and pressure field evolutions as a function of microstructural morphology [6]. However, the previous work was limited to a single shock loading condition. Because it is required to analyze the SDT of EM subjected to various shock strengths, here we extend the original PARC architecture to allow it to make predictions with arbitrary input shock strengths.

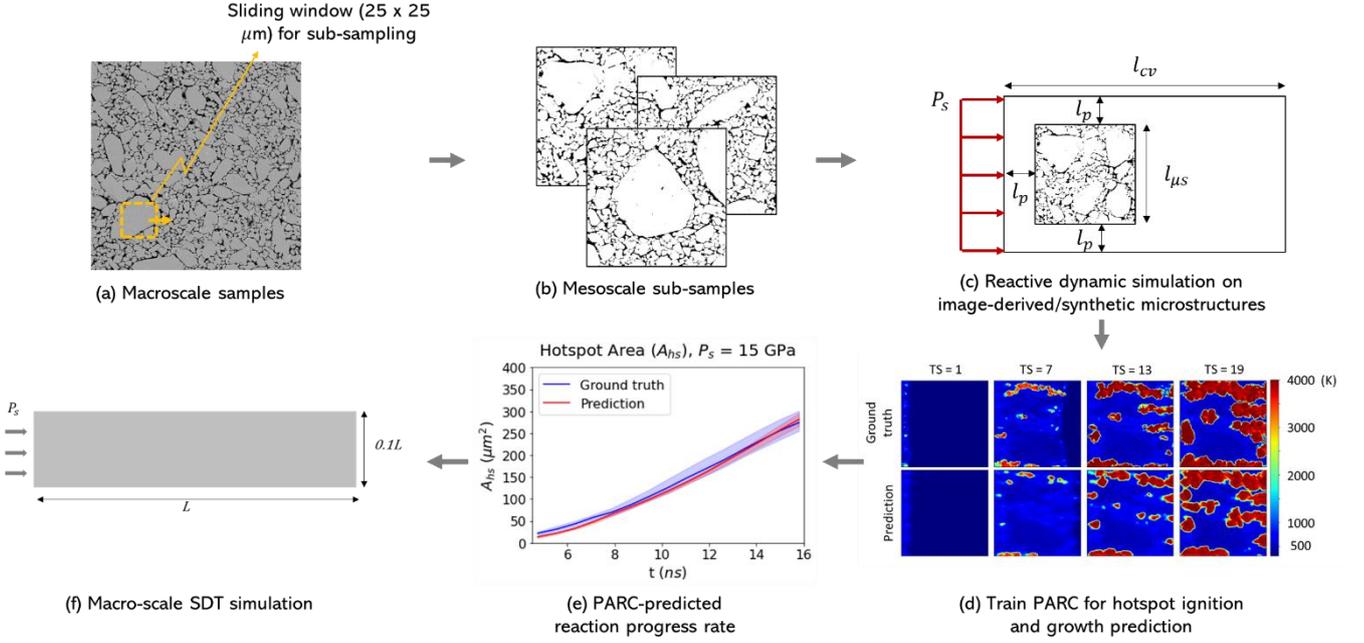

**Figure 1:** Overview of the proposed multiscale SDT simulation framework. From an image-derived macrosample (a), mesoscale sub-samples (b) of HMX microstructures are collected. Consequently, the reactive dynamic simulations on several mesoscale sub-samples (c) are performed and the resulting temperature and pressure field evolutions are used to train PARC. After training, PARC is used to replace DNS to compute the temperature field evolution of shocked HMX microstructures under different shock strengths (d). From PARC-predicted temperature field evolutions, reaction progress rate can be derived (e) and used to inform the macroscale SDT simulation (f) to measure the distance-to-detonation of EM under an applied shock.

It is well known that the reaction rate of EM is a function of the microstructural morphology and the applied shock pressure [5]. In a uniform Eulerian grid, at a given position $r(x,y)$ and time $t$, the differential governing equation describing the rate of change of state variables $X(r;t)$ (temperature and pressure), can be formulated as:

$$\frac{\partial X}{\partial t} = f(X, \mu, P_s)$$
$$X(r; t=0) = [T_0(r), P_0(r)] \tag{1}$$

In Eq. (1), $\mu$ is the shape descriptor, introduced to account for the effect of the microstructural morphology in the thermo-mechanics of shocked EM. Here, $P_s$ is an additional parameter that was not included in the original formulation of PARC [3] and is introduced into the architecture to account for different shock loading conditions.

Note that the explicit differential equation governing temperature $T(r;t)$ and pressure $P(r;t)$ is unavailable. Instead, the evolution of temperature and pressure fields within EM microstructures in response to shock are components of a larger system of equations that describes the compressible multimaterial reaction dynamics hyperbolic conservation laws associated with material deformation and microstructure evolution [19,26,27]. Due to the complexity of the reactive dynamics flow system, it is infeasible to incorporate the full system into PARC neural network architecture. Therefore, the differential equation modeling the energy localization in shocked EM and its solution with respect to two state variables $(T, P)$ were both learned from data.

Eq. (1) is a partial differential equation with given initial conditions, which can be solved by integration with a time interval $\Delta t$. The state $X$ at a given time $t$ can be computed as:

$$X(r; t + \Delta t) = X(r; t) + \int_{t}^{t+\Delta t} f(X(r; t), \mu, P_s)\, dt, \tag{2}$$

Nguyen et al. [6] introduced a deep learning approach for solving the above equation. First, the function $f$ in Eq. (1) is modeled as a CNN (named "differentiator"), based on the universal approximation theorem that CNNs with appropriate width (i.e., number of neurons/channels) and depth (i.e., number of layers) can represent an unknown function [29]:

$$f(X, \mu, P_s \mid \theta) := f(X, \mu, P_s), \tag{3}$$

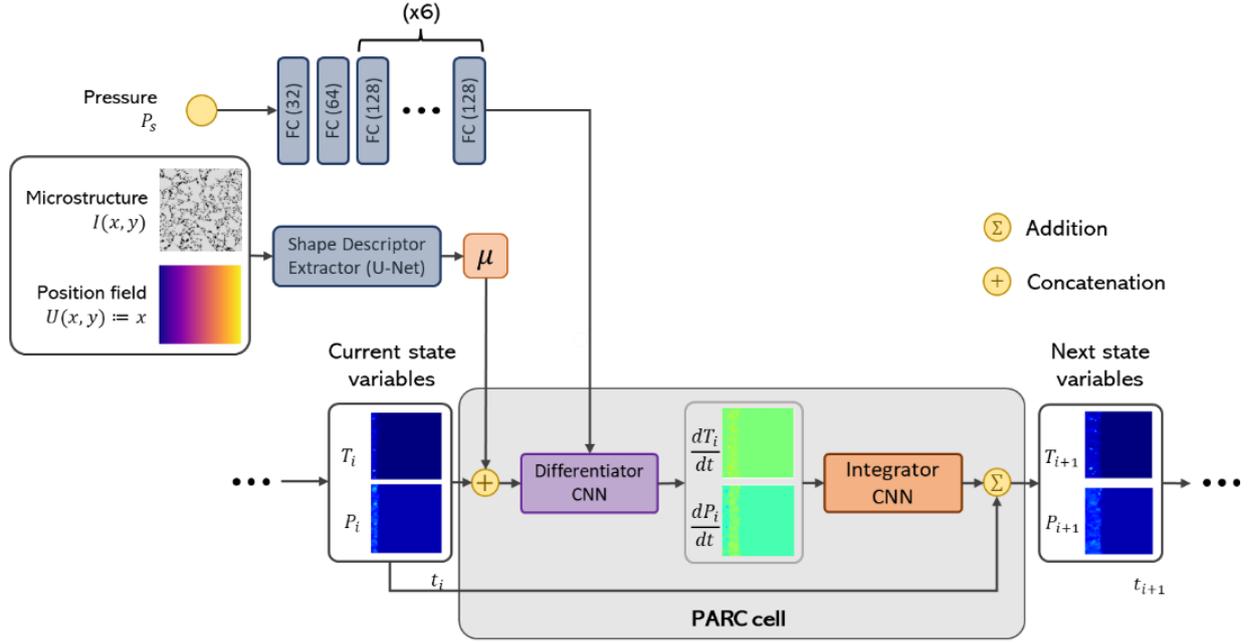

**Figure 1** Overall architecture of the enhanced PARC for hotspot thermo-mechanics modeling with different given applied shock pressures. To account for the influence of the applied shock, the applied pressure value is first transformed into a 128-feature vector using a fully connected neural network. At each time step, the differentiator CNN takes the shape descriptor $\mu$, the pressure feature vector $P_f$, and the temperature and pressure fields of the previous time step $X(r; t_{i-1})$, as inputs to derive time derivative fields of the temperature and pressure, $\dot{X}(r; t_i)$. The temperature and pressure time derivative fields are used as input for the integrator CNN to compute the actual change in temperature and pressure. The outputs of the integrator CNN are then added to the temperature and pressure fields of the previous time step to derive the temperature and pressure fields of the next time step.

where $\theta$ denotes the neural network parameters (weights and biases) of the differentiator CNN.

Second, in PARC, the differential equation is solved using a data-driven method, with the integral in Eq. (2) being approximated with another CNN (named "integrator").

$$S(f|\varphi) = \int_{t_{i-1}}^{t_i} f(X, \mu, P_s) \, dt, \tag{4}$$

where $\varphi$ is neural network parameters of the integrator CNN. The introduction of the data-driven integration is motivated by the hotspot thermo-mechanics of shocked EM microstructures which is highly transient and nonlinear; therefore, approaches using naïve numerical integrations with fixed time step values may accumulate errors and decrease the accuracy of the prediction. Although adaptive numerical integration solvers can help overcome this issue, their computation costs are high when the nonlinearity of the system increases. Instead, the efficiency of data-driven integration is exploited in this work.

Using such deep learning formulations, the state prediction problem in Eq. (2) is expressed as:

$$X(r; t_i) = X(r; t_{i-1}) + S(f(X, \mu, P_s \mid \theta) \mid \varphi), \tag{5}$$

where $t_i$ is a discrete time step. Figure 2 is a graphical illustration of the above mathematical procedure. At a given time step $t_i$, the U-Net [30] is used to encode the microstructure image, $I(x, y)$ and the corresponding position map $U(r): (x, y) \mapsto x$ into the shape descriptor $\mu$ of dimension $(M \times N) \times 128$, where $M \times N$ is the spatial dimension of the microstructure image. Meanwhile, the applied pressure, $P_s$, is transformed into a feature vector, $P_f$, using an artificial neural network consisting of a stack of fully connected (FC) layers [31]. The pressure transformation network includes one FC layer with 32 neurons, one FC layer with 64 neurons and six FC layers with 128 neurons. This network design results in a feature vector $P_f$ with 128 features. The differentiator CNN takes the shape descriptor $\mu$, the pressure feature vector $P_f$, and the temperature and pressure fields of the previous time step $X(r; t_{i-1})$, as the input to derive time derivative fields of the temperature and pressure, $\dot{X}(r; t_i)$. Consequently, the computed time derivatives are sent to the integrator CNN for the calculation of time integration $\int_{t_{i-1}}^{t_i} \dot{X} \, dt$. The calculated values are added to the state values of the previous time step $\dot{X}(r; t_{i-1})$ to derive the state values of the current time step $\dot{X}(r; t_i)$. The process is continued recursively until the final time step is reached. Within the computational process, the network parameters of the differentiator CNN (purple box) and the integrator CNN (orange box) are shared between time steps, making the architecture recurrent. Compared to the original PARC architecture, the current enhanced model is capable of accounting for the effects of different shock strengths as the differentiator CNN takes the applied pressure value as one of its inputs to compute the time derivatives.

*3.2.2 Training Objectives*

The goal of PARC is to learn the governing differential equations, i.e., accurately predict the time derivatives of temperature and pressure fields and solve the differential equations via data-driven integration, i.e., accurately predict the temperature and pressure fields in the next time steps. Therefore, the training of PARC can be cast as an optimization problem in which the loss function as described below is minimized with respect to the neural network parameters $\theta$ and $\varphi$:

$$L(\theta, \varphi \mid \hat{X}) = \int_0^\infty \left\| \hat{X}(r; \tau) - X_{i-1} - S(f(X, \mu \mid \theta) \mid \varphi) \right\|_2 d\tau + \int_0^\infty \left\| \hat{\dot{X}}(r; t) - f(X, \mu \mid \theta) \right\|_2 d\tau, \tag{6}$$

where the hat (^) denotes the ground truth data, derived from DNS simulation results. The continuous temporal domain is tessellated with reasonably small time intervals $\Delta t$; therefore, Eq. (7) can be rewritten in the discrete form:

$$L(\theta, \varphi \mid \hat{X}) = \sum_{t_i} \left\| \hat{X}(r; t) - X_{i-1} - S(f(X, \mu \mid \theta) \mid \varphi) \right\|_2 + \sum_{t_i} \left\| \hat{\dot{X}}(r; t) - f(X, \mu \mid \theta) \right\|_2 \tag{7}$$

*3.2.3 Data and training*

The enhanced PARC model was trained with 40 simulation instances on microstructures of HMX, a typical type of EM. All the microstructures are acquired from SEM with the dimension of 25 ×25 μm, resolved in 240×240-pixel images. SCIMITAR3D [19,26–28], a multi-material reactive dynamic code, was used to compute the temperature and pressure field evolutions of HMX microstructures under three different shock strengths: $P_s = 5, 9.5$, and 15 (GPa). The DNS-computed temperature and pressure field evolutions, including snapshots of temperature and pressure fields of 19 discrete time steps with equal time intervals $\Delta t = 1.09, 0.79, 0.72$(ns), for $P_s = 5, 9.5$, and 15 (GPa), respectively. The temperature and pressure fields were also given in a 240×240 pixels image format. There was a difference in the time interval value of different shock strengths due to the difference in the total simulation time for different shock strengths and the requirement from PARC to maintain the total number of discrete time steps. Among 40 simulation instances used for training, 30 of them have $P_s = 9.5$ (GPa), five of them have $P_s = 5$ (GPa), and five of them have $P_s = 15$ (GPa).

Our training procedure is divided into two stages. In the first stage, we trained the model for $P_s = 9.5$ (GPa) using all 30 data samples for 500 epochs. Consequently, we added the data for $P_s = 5$ & 15 (GPa), excluded the shape descriptor network parameters from the training, and continued the training for another 500 epochs. Here we hypothesize that with the physics-aware architecture, PARC can be generalized from training result of one shock strength to make predictions on other shock strengths with reasonable amount of training data.

For the testing set, three simulation instances that had not appeared in the training set were included for each shock strength. All the data used for the training was normalized to have values ranging from -1 to 1. The neural networks in the model were initialized with the normalized He initialization method [32] and the ADAM optimizer [33] with a learning rate of $5 \times 10^{-5}$ was used to train the model.

## 3.3 Computing reaction progress rates with PARC

The predictions made by PARC from Section 3.2 can be utilized to quantify the intensity and contribution of the energy localization due to the void collapse. The critical hotspots resulting from the void collapse are identified based on temperature field evolutions predicted by PARC. Particularly, a constant temperature threshold of 875 K is used to delineate the border between critical hotspots and the bulk of non-reacting/unreacted material. Furthermore, for simplification, we assume that the final species mass fraction $Y_n = 1$ inside hotspots (fully gas products) and $Y_n = 0$ outside hotspots (i.e., in the unreacted material). For the Tarver-3 step chemical reaction model [4,20], we set $n = 4$.

The state of the reacting mixture is determined by the reaction progress variable $0 \leq \lambda \leq 1$, defined as the mass fraction of the reaction product $M_{product}$ to the initial total control volume mass $M_{cv}$, i.e.,

$$\lambda = \frac{M_{product}}{M_{cv}}, \tag{8}$$

where, the total mass of material $M_{cv}$ in the control volume is computed as:

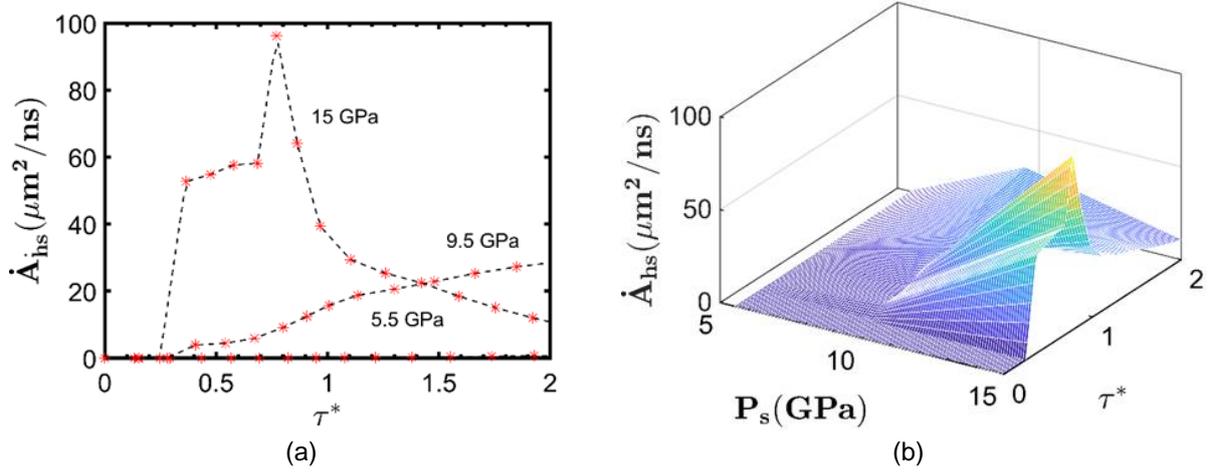

**Figure 2** (a) Sample averaged $\dot{A}_{hs}$ computed by PARC for three values of $P_s$ and $\tau^*$ ranges from 0 to 2. Dased lines indicate results obtained by piecewise interpolation from PARC predicted values (dots). (b) The hypersurfaces $\dot{A}_{hs}(P_s, \tau^*)$ constructed using linear interpolation using PARC predicted values.

$$M_{cv} = \rho_{HMX} A_{cv}. \tag{9}$$

In Eq. (9), $\rho_{HMX} = 1905 \ kg/m^3$ and $A_{cv}$ is the total volume (area in 2D case) of the control volume. The total mass of reacted gas products accumulated in the control volume $\Omega$ is computed as:

$$M_{product} = \int_\Omega \rho \, Y_n \, dA. \tag{10}$$

Since $Y_n = 1$ inside the hotspots (T>875 K) and 0 elsewhere, Eq. (10) becomes:

$$M_{product} = \int_{\Omega_{hs}} \rho \, dA_{hs}. \tag{11}$$

In addition, the density of gaseous material inside hotspots is considered having a uniform value, $\rho_{product} = 2300 \frac{kg}{m^3}$, thus $M_{product} = \rho_{product} A_{hs}$. Therefore, the reaction progress rate can be computed as:

$$\lambda = \frac{\rho_{product}}{\rho_{HMX}} \frac{A_{hs}}{A_{cv}} \tag{12}$$

or in its rate form:

$$\dot{\lambda} = \beta \dot{A}_{hs}, \tag{13}$$

where $\beta = \frac{\rho_{reacted}}{\rho_{HMX}} \frac{1}{A_{cv}}$. To model completion of the reaction as $\lambda \to 1$, the form factor $S(\lambda) = 1 - \lambda$ is introduced on the right hand side of Eq. (13) to account for the effect of decaying reaction rate as the material gradually reaches final state:

$$\dot{\lambda} = \beta \dot{A}_{hs}(1 - \lambda). \tag{14}$$

Therefore, the reaction progress rate $\dot{\lambda}$ is obtained as a function of the hotspot area rate of change $\dot{A}_{hs}$ which is derived from the $A_{hs}$ vs $t$ curve shown in Fig. 1e. $\dot{\lambda}$ is used in the macroscale simulation to close the governing equation system. The derivation of the $A_{hs}(t)$ function is described as follows.

At a time $t_k$, the PARC-predicted temperature field of a control volume $\boldsymbol{T}(t_k)$ is tessellated onto a $M \times N$ uniform grid. The total hotspot area $A_{hs}$ is computed based on the temperature field as:

$$A_{hs}(t_k) = \sum_{i=1}^{M} \sum_{j=1}^{N} A_{ij}^{hs}(t_k), \tag{15}$$

with:

$$A_{ij}^{hs}(t_k) = \begin{cases} A_{ij}(t_k) & if\ T_{ij}(t_k) \geq 875K \\ 0 & if\ T_{ij}(t_k) < 875K \end{cases}. \tag{16}$$

In Eqs. (15)-(16), $T_{ij}$ is the temperature value at the $(i,j)$-th grid location, $A_{ij}^{hs}$ is the area of the grid cell $(i,j)$-th occupying the hotspot region, and $A_{ij}$ is the area of a single grid location $(i,j)$-th, which is a uniform value given by:

$$A_{ij} = \frac{l^2}{MN}, \tag{17}$$

where $l$ is the spatial dimension of the control volume, in this case, $25\ \mu m$. For a given $P_s$ value, multiple samples are used to perform the simulation and calculate $A_{hs}$. Consequently, a sample-averaged $A_{hs}$ is used to compute the reaction progress rate $\lambda$. We sampled 100 mesoscale sub-samples with dimension of $25 \times 25\ (\mu m)$ for the computation of $A_{hs}$. Note that all sub-samples are Class V HMX, which has been reported to be more homogeneous compared to other classes of HMX [34] and consequently have low variability in reaction progress rate [16].

PARC predicted reaction progress rate is used to inform the macro-scale solver using a surrogate model, similar to our previous work [5]. The scale-bridging surrogate model was trained with sampled-averaged $\dot{A}_{hs}$ for each applied shock strength computed by PARC. For a control volume, $\dot{A}_{hs}$ is a function of local shock pressure $P_s$ and time. The $\dot{A}_{hs}$ surrogate is constructed in the following form:

$$\dot{A}_{hs} = f(P_s, \tau^*). \tag{18}$$

In Eq. (18), $\tau^*$ is the local control volume time, normalized by the shock time scale and defined as: $\tau^* = (t - t_{shift})/(l/U_s)$, where $t_{shift}$ is time that shock reaches the control volume and $U_s$ is the shock velocity in solid HMX. In this work, because of the high computational cost, the surrogate is constructed from just three values of $P_s$, which are 5.5, 9.5 and 15 GPa (Fig. 3(a)). As indicated by Rai et al. [26], for low $P_s$, no hotspots could develop as they are quenched by diffusion. Therefore, in the macroscale simulation, $\dot{A}_{hs}$ corresponding to $P_s$ smaller than pressure lower limit, 5 GPa, is set to zero. The upper limit of $P_s$ is taken to be 60 GPa, which is the estimated value of the Von Neumann spike pressure by Massoni et al. [14]. The values of $\dot{A}_{hs}$ subjected to $P_s$ varying from 5.5 GPa to 15 GPa are obtained by piecewise linear interpolation. The $\dot{A}_{hs}(P_s, \tau^*)$ hypersurface is as depicted in Fig. 3(b).

## 3.4 Macroscale SDT simulation

At the macroscale, as mentioned above, the EM is treated as homogenous and the behavior of the material at macroscale during the SDT process is described using the system of equations:

$$\boldsymbol{D}(\boldsymbol{a}|\lambda) = 0. \tag{19}$$

In the above Eq. (19), $\boldsymbol{D}$ represents the set of hyperbolic conservation laws and constitutive relations of the homogenized material; details on the equations and solution procedures were provided in several previous works [4,5]. The vector $\boldsymbol{a}$ denotes the set of variables of the thermo-mechanical flow, including velocity components, the density, and the internal energy of the homogenized mixture. Finally, the reaction process variable $\lambda$, is used to define the progress of reaction, calculated by the mass fraction of reaction products in a given macroscale control volume. As the imposed shock travels through the examined domain, the SDT transition is distinguished by the conversion of cold unreacted solid to the reacted product mixture, following the Rayleigh line [35]. This transition is dictated by the value of $\lambda$ and the evolution equation for $\lambda$ can be derived from analysis results at the mesoscale [4,5,16,18]. In the current work, the evolution equation of $\lambda$ is derived from the predictions made by PARC.

The overall procedure for the meso-informed SDT simulation could be summarized as follows:
1) The trained PARC is used to predict the temperature and pressure field evolutions of different HMX microstructures, subjected to different $P_s$.
2) Consequently, the evolution of $A_{hs}$ subjected to different shock strengths are derived. For the sake of simplicity, the averaged value of $A_{hs}$ across multiple HMX microstructures is used in this work.
3) From the derived $A_{hs}$ evolution, the surrogate model for $\dot{A}_{hs}$ subjected to different shock strengths is constructed in the form given by Eq. (18)
4) The governing equation, including conservation of mass, momentum and energy, the equation of state (Jones-Wilkins-Lee is used in this work) are resolved numerically for which details could be found from [5].
5) The local time and local pressure at each control volume are used to probe the value of $\dot{A}_{hs}$ (Eq. (18)) which is then converted to $\dot{\lambda}$ using Eq. (14). $\dot{\lambda}$ is used to update the value of $\lambda$ using Runge-Kutta-Fehlberg method [36] and Strang operator splitting method [37].

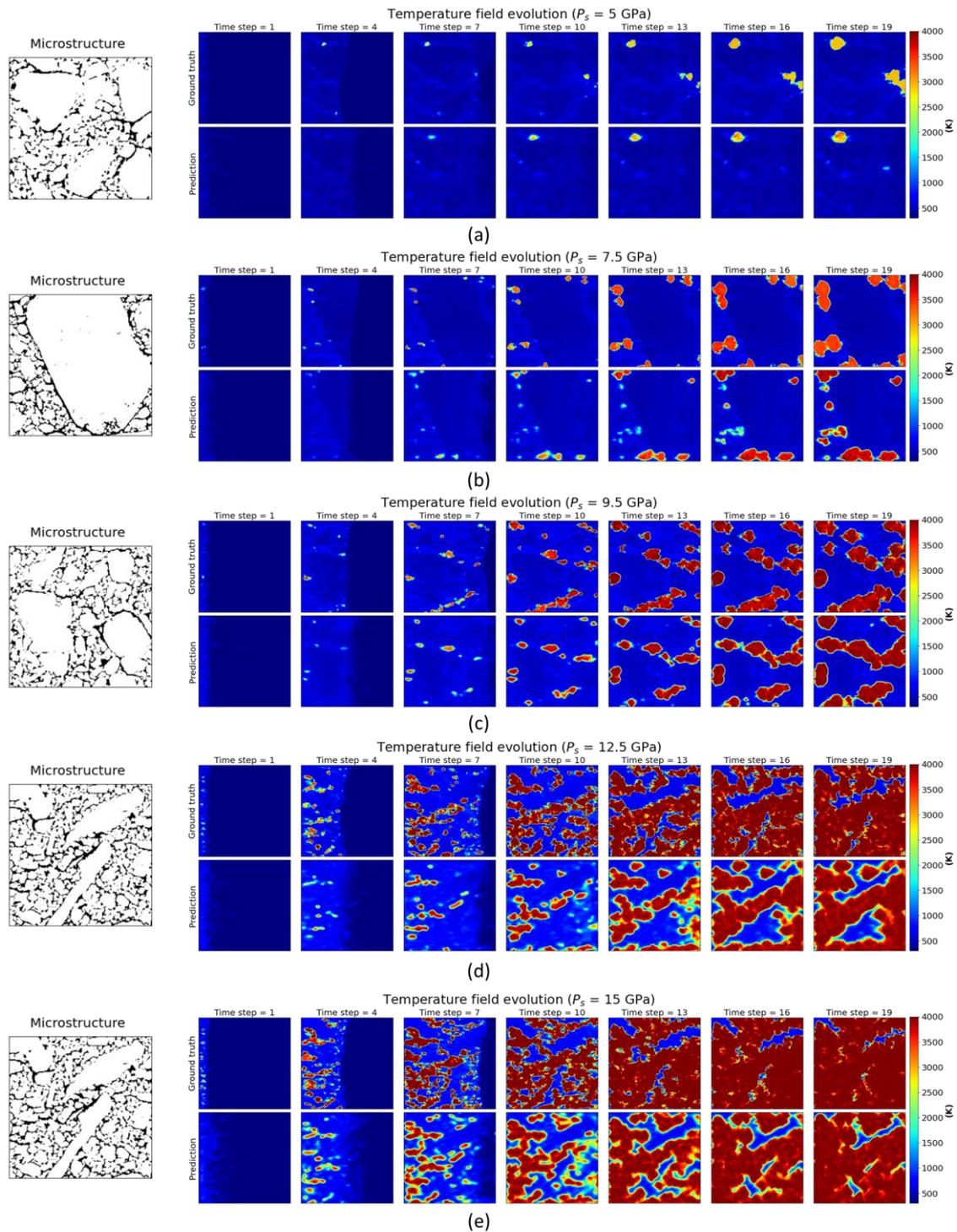

**Figure 4:** The temperature field evolutions predicted by PARC under different applied shock pressure and their corresponding microstructure. (a) $P_s$ = 5 GPa. (b) $P_s$ = 7.5 GPa. (c) $P_s$ = 9.5 GPa. (d) $P_s$ = 12.5 GPa. (e) $P_s$ = 15 GPa. As can be seen, PARC-predicted temperature field evolutions agree well with their corresponding ground truth computed by DNS even with the shock pressure value not seen during training.

### 3.5 Measuring the performance of EM from macroscale computations

In this work, the performance of EM can be measured by calculating the critical energy [38–41] demarcated graphically by a James envelope [38]. The critical energy is a function of the input shock pressure $P_s$ and shock pulse duration $\tau_s$ [41]. In the $P_s - \tau_s$ space, the James envelope delineating the go-no-go subspaces is obtained; for points above the James curve SDT occurs while below the James curve the impact is subcritical, i.e., SDT fails to occur. In this work, following earlier experiments [41], the criticality of EMs was determined to lie along a curve defined by $P_s^2 \cdot \tau_s = constant$ curve [34].

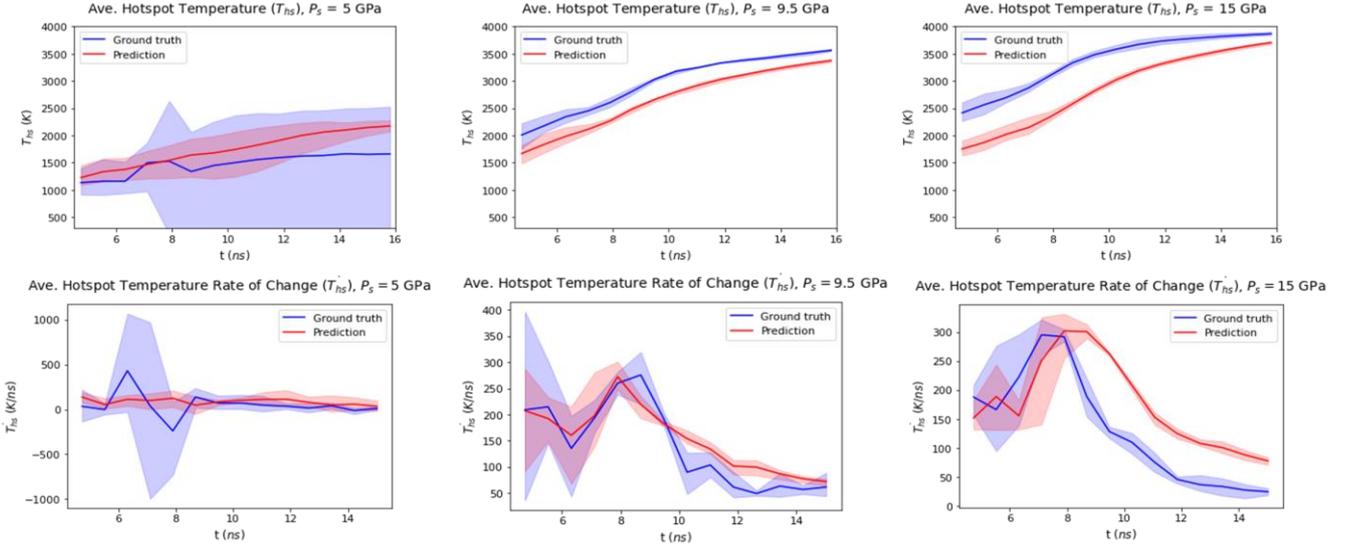

**Figure 5:** Validation of PARC-predicted average hotspot temperature (top) and its rate of change over time (bottom). The applied pressure (from left to right) are: $P_s$ = 5, 9.5, and 15 GPa. For all three tested applied shock pressure, there is an agreement between PARC prediction and the corresponding ground truth.

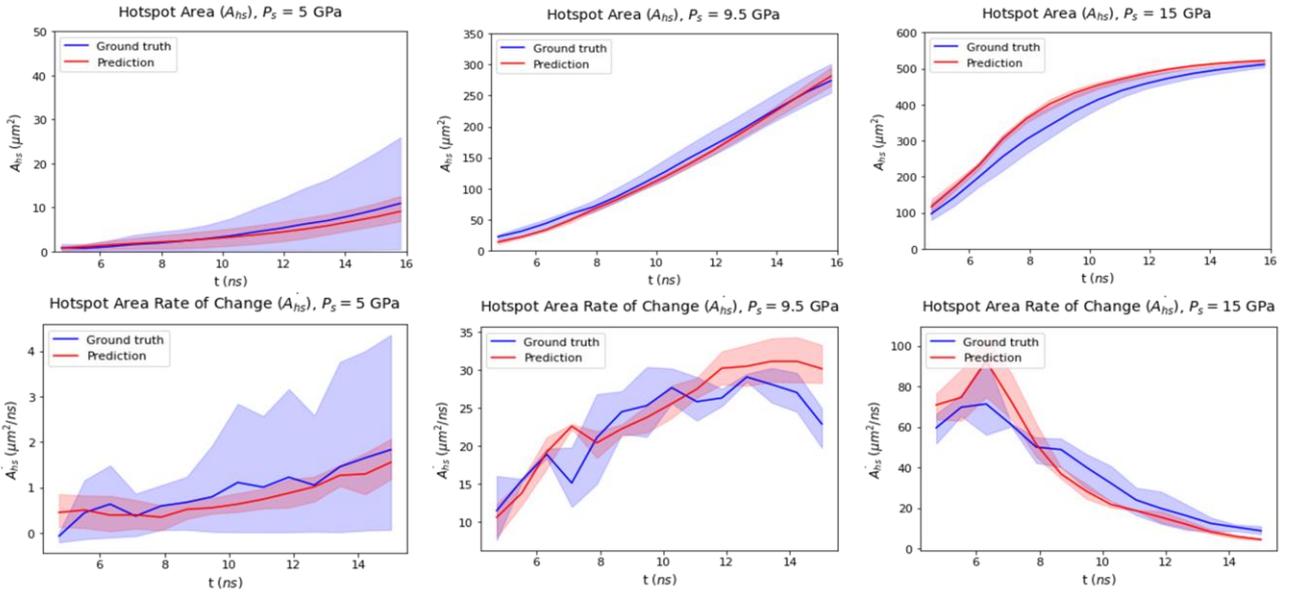

**Figure 6:** Validation of PARC-predicted total hotspot area (top) and its rate of change over time (bottom). The applied shock strengths (from left to right) are: $P_s = 5, 9.5,$ and $15$ GPa. Again, for all three tested applied shock pressure, there is an agreement between PARC prediction and the corresponding ground truth.

## 4 Results and discussion

### 4.1 Mesoscale hotspot physics prediction accuracy under different shock strengths

Figure 4 illustrates the temperature field evolution predictions by PARC and the corresponding DNS ground truth for five different shock strengths: $P_s = 5, 7.5, 9.5, 12.5$ and $15$ (GPa). Qualitatively, the temperature field evolution predicted by PARC agrees well with those from DNS predictions even at two unseen shock pressures, viz. $P_s = 7.5$ and $12.5$ (GPa). In addition, hotspot locations are well recognized across multiple different shock strengths. Moreover, the propagation of shock waves entering EM microstructures is also well modeled.

In addition to the above qualitative evaluation, we also quantitatively validated the enhanced PARC model using sensitivity metrics as proposed by Nguyen et al. [6], including average hotspot temperature and total hotspot area as well as their rate of change over time. Fig. 5 and Fig. 6 show the validation results of PARC with EM sensitivity metrics. As illustrated, sensitivity metrics derived from PARC predicted temperature field evolutions agree with the ones derived from DNS. Moreover, the prediction root-mean-square-error (RMSE) w.r.t DNS ground truth value across all 9 test instances are relatively low with the average value of 763 (K), 25.7 ($\mu m^2$), 159.08 (K/ns) and 7.9 ($\mu m^2/ns$) for average hotspot temperatures, total hotspot area, and their rate of changes over time, respectively.

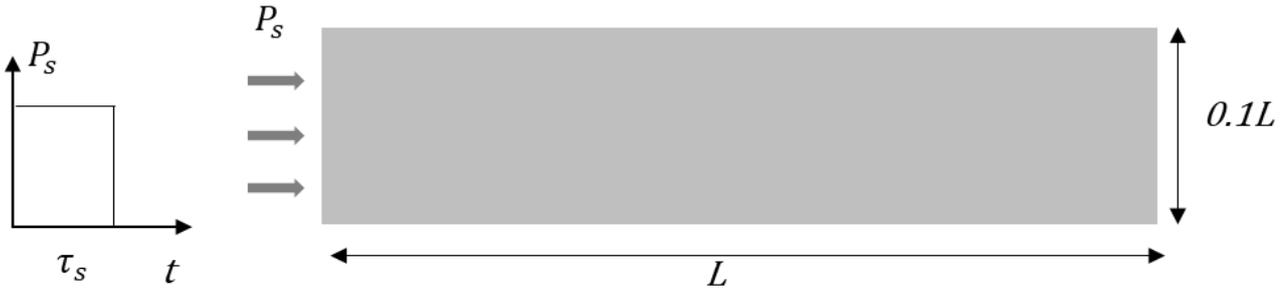

**Figure 7:** Macro SDT simulation set up. A homogenized material is subjected to a shock pulse of pressure $P_s$ for a duration $\tau_s$ from the west side of the boundaries. The sample width L is 1 mm and the sample height is 0.1 L. The macro simulation is resolved by a uniform grid with grid size $\Delta x = 2\ \mu m$.

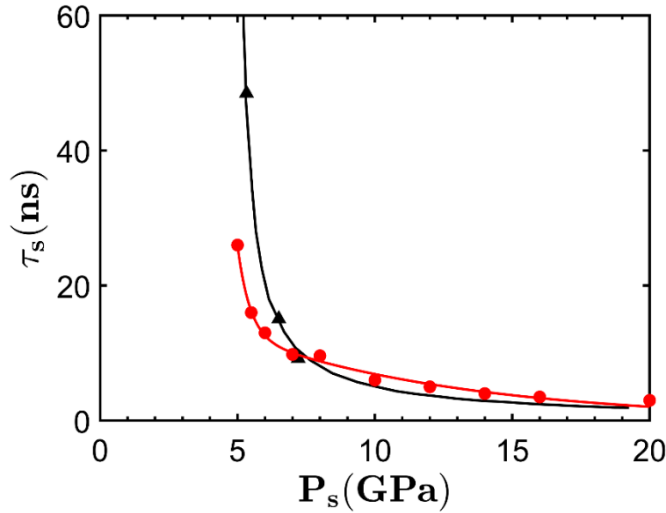

**Figure 8:** James criticality envelop for Class V HMX material from experiment (black) [34] and PARC prediction (red). The solid lines are $P_s^2 \tau_s = constant$ curves fitted to data points.

From the above validation results, PARC is accurate enough to replace DNS in the multiscale simulation framework. In addition, as being demonstrated by Nguyen et al. [6], the computation cost of PARC is by multiple order lower than the one of DNS. These added benefits of PARC would help facilitate the distance-to-detonation estimation at macroscale and will accelerate the process of designing new EM, which will be showed in the following section.

## 4.2 SDT simulation using PARC

The setup for the macroscale computation is as illustrated in Fig. 7. A rectangular shock pulse of pressure $P_s$ is applied for a duration $\tau_s$ from the left of the domain. In addition, a uniform grid space $\Delta x = 2\ \mu m$ that has been tested for grid convergence is used. For each shock input $P_s$ value varying from 5 to 20 GPa, the critical shock pulse duration $\tau_{critical}$ is computed and plotted against $P_s$ as illustrated in Fig. 8. The simulation result was compared with experimental data for Class V HMX obtained from the James's curve as reported by Welle et al. [34]. As shown in Fig. 8, there is a well agreement between the simulation and the experiment, showing that the PARC-based $\tau_{critical}$ computation is accurate. Such an agreement proves that PARC can be used in a multiscale SDT simulation framework and can contribute to the reduction the multiscale simulation time with enhanced the sub-grid physics representation.

# 5 Conclusion

This work presented a deep learning approach for multiscale SDT simulation of heterogeneous EM. The presented deep-learning model, PARC, could provide rapid and accurate predictions of the hotspot evolution within shocked EM microstructures. Based on PARC-predicted temperature evolution fields, the reaction progress rate could be derived and used to inform the macroscale SDT simulation. The distance-to-detonation estimated by our multiscale framework was validated with experimental James' curve, indicating a well agreement between the experiment and the simulation. With these enhanced predictive capabilities enabled by PARC, SDT simulations of EM can be facilitated, and the efforts required to design new EM can be reduced.

Future work will be directed toward enhancing the accuracy of the PARC predicted temperature field evolution. Despite providing QoI predictions that agree well with ground truth DNS, the point-wise accuracy of PARC predicted temperature field evolutions remains to be examined and improved. To this end, we are currently working on developing more rigorous continuous convolutional kernels to improve the point-wise accuracy of PARC-predicted field evolutions. In addition, our future work will extend the modeling capability of PARC to learn the hotspot physics of other EM species as well as other types of loading conditions, e.g., applied shock with pulse duration.

*Acknowledgements*

This material is based upon work supported by the National Science Foundation under Grant No. 2203580 and by the U.S. Air Force Office of Scientific Research (AFOSR) Multidisciplinary University Research Initiative (MURI) program (Grant No. FA9550-19-1-0318; PM: Dr. Chiping Li, Dynamic Materials program).

# Appendix

## Sensitivity metric computation

The sensitivity QoIs of EM—the average hotspot temperature, $\overline{T^{hs}}$, total hotspot area, and their rate of change over time, a time $t_k$—were computed as:

$$\overline{T^{hs}}(t_k) = \frac{\sum_{i=1}^{M}\sum_{j=1}^{N}\left(T_{ij}^{hs}(t_k)A_{ij}^{hs}(t_k)\right)}{A^{hs}(t_k)}, \tag{A1}$$

$$T_{ij}^{hs}(t_k) = \begin{cases} T_{ij}(t_k) \text{ if } T_{ij}(t_k) \geq 875\ K \\ 0 \text{ if } T_{ij}(t_k) < 875\ K \end{cases}, \tag{A2}$$

$$A_{hs}(t_k) = \sum_{i=1}^{M}\sum_{j=1}^{N} A_{ij}^{hs}(t_k), \tag{A3}$$

$$\dot{\overline{T^{hs}}}(t_k) = \frac{\overline{T^{hs}}(t_k) - \overline{T^{hs}}(t_{k-1})}{t_k - t_{k-1}}, \tag{A4}$$

$$\dot{A}^{hs}(t_k) = \frac{A^{hs}(t_k) - A^{hs}(t_{k-1})}{t_k - t_{k-1}}. \tag{A5}$$

In Eqs. (A1-A5), $T_{ij}^{hs}$ and $T_{ij}$ are the hotspot temperature and the temperature values at the $(i,j)$-th grid location. $A_{ij}^{hs}$ is the area of a grid cell in the hotspot region. Since a uniform grid is applied, $A_{ij}^{hs}$ is a constant value.

## Evaluation metrics for sensitivity prediction

In this work, the root-mean-squared error was used to evaluate the sensitivity prediction of PARC:

$$RSME = \sqrt{\frac{\sum_{l=0}^{Q}\sum_{k=0}^{P}[\hat{X}_l(t_k) - X_l(t_k)]^2}{QP}}. \tag{A6}$$

Here, $\hat{X}_l(t_k)$ and $X_l(t_k)$ are the ground truth and sensitivity QoI predicted by PARC of $l$-th testing sample at time $t_k$. Further, $P$ is the number of total time step and $Q$ is the number of testing samples